\documentclass[twocolumn,showpacs,preprintnumbers,amsmath,aps,prb]{revtex4-2}
\usepackage{graphicx}
\usepackage{dcolumn}
\usepackage{bm}
\usepackage{verbatim} 
\usepackage[usenames, dvipsnames]{color} 
\usepackage{ulem}
\usepackage{color}
\usepackage[colorlinks=false,pdfencoding=auto,psdextra]{hyperref}
\usepackage{multirow}
\usepackage{tabularx}
\usepackage{dsfont}
\usepackage{soul}

\definecolor{orange}{rgb}{1.0, 0.5, 0.0}

\newcommand{\CFI}{FA$_{0.95}$Cs$_{0.05}$PbI$_3$ }
\newcommand{\MFI}{MA$_{0.4}$FA$_{0.6}$PbI$_3$ }

\newcommand{\addV}[1]{\textcolor{red}{ #1}}

\begin{document}

\title{Millisecond spin coherence of electrons in semiconducting perovskites revealed by spin mode locking
}

\author{Sergey R. Meliakov$^{1}$, Evgeny A. Zhukov$^{1}$, Vasilii V. Belykh$^{1}$, Dmitri R. Yakovlev$^{1}$, Bekir Turedi$^{2,3}$, Maksym V. Kovalenko$^{2,3}$, and Manfred Bayer$^{1,4}$} 

\affiliation{$^{1}$Experimentelle Physik 2, Technische Universit\"at Dortmund, 44227 Dortmund, Germany}
\affiliation{$^{2}$Department of Chemistry and Applied Biosciences,
Laboratory of Inorganic Chemistry, ETH Z\"{u}rich, 8093 Z\"{u}rich, Switzerland}
\affiliation{$^{3}$Department of Advanced Materials and
Surfaces, Laboratory for Thin Films and Photovoltaics, Empa - Swiss Federal Laboratories for Materials Science and Technology, 8600 D\"{u}bendorf, Switzerland}
\affiliation{$^4$Research Center FEMS, Technische Universit\"at Dortmund, 44227 Dortmund, Germany}

\date{\today}

\begin{abstract}

Long spin coherence times of carriers are essential for implementing quantum technologies using semiconductor devices for which, however, a possible obstacle is spin relaxation. For the spin dynamics, decisive features are the band structure, crystal symmetry, and quantum confinement. Perovskite semiconductors recently have come into focus of studies of their spin states, notivated by efficient optical access and potentially long-living coherence. Here, we report an electron spin coherence time $T_2$ of the order of 1~ms, measured for a bulk \CFI lead halide perovskite crystal. Using periodic laser pulses, we synchronize the electron spin Larmor precession about an external magnetic field in an inhomogeneous ensemble, the effect known as spin mode locking. It appears as a decay of the optically created ensemble spin polarization within the dephasing time $T_2^*$ of up to 20~ns and its revival during the spin coherence time $T_2$ reaching the millisecond range. This exceptionally long spin coherence time in a bulk crystal is complemented by millisecond-long longitudinal spin relaxation times $T_1$ for electrons and holes, measured by optically-detected magnetic resonance. These long-lasting spin dynamics highlight perovskites as promising platform for the quantum devices with all-optical control. 

\end{abstract}

\maketitle




The coherent spin dynamics of electrons and holes in semiconductor nanostructures have been extensively studied over the past  decades~\cite{Henneberger2009,Benyoucef2023,Spin_book2017}. Generation, detection, and manipulation of spin polarization by laser excitation have been demonstrated~\cite{Yakovlev_Ch6,Greilich2009_natp}, opening pathways towards spintronic devices, which can operate down to the quantum level. 
The spin coherence time $T_2$ of a charge carrier is a key parameter for quantum technology applications, 
which should be maximized, e.g., for storing quantum information. For processing it, the coherence time has to be compared to the time needed for manipulating the associated quantum state, which is in the nanosecond range using microwave pulses and in the picosecond range or even shorter using optical pulses. 
Different materials have been examined for achieving long spin coherence times. $T_2$ has been found to be limited to not more than microseconds 
for band states in semiconductor nanostructures in a cryogenic environment. This time likely is too short for applications, in particular in comparison to other platforms such as atoms or ions offering coherences lasting seconds. For superconductors, requiring microwave quantum state manipulation, $T_2$ has improved to about milliseconds over the years. Targeting spin coherences lasting milliseconds also for semiconductors has remained a goal which might make them competitive with superconductors, in particular, in conjunction with optical control tools that exploit the large oscillator strength of band-to-band transitions.

In fact, spin coherences approaching milliseconds have been demonstrated for strongly localized crystal defect states such as NV centers in diamonds, divacancies in SiC, or phosphorous centers in Si at temperatures in the range of $1.6-4.2$~K, see for a review, e.g., Ref.~\onlinecite{Stano2022} and Table~\ref{tab:T2_times}. These coherence times might be increased further by implementing protocols such as dynamic decoupling from surrounding baths using elaborated pulse sequences, further temperature reduction to milliKelvins to calm these baths, isotopic purification of materials to avoid nuclei with non-zero spin, etc. However, not all of these materials are well suited for optical spin processing, which is considered an important factor to achieve the required number of manipulations. Thus, there is still a high interest for semiconducting materials combining superior spin coherence with efficient optical access.



Lead halide perovskite semiconductors recently have emerged as promising material system for spin physics~\cite{Wang2019,Ning2020,Kim2018,Wu2024Review}. They have a simple band structure~\cite{Kovalenko2017} (spin 1/2 states at the bottom of conduction band and top of valence band) and show spatial inversion symmetry resulting in suppression of one of the strongest spin relaxation mechanisms in semiconductors, the Dyakonov-Perel mechanism~\cite{Kopteva2024OO}. Further, they exhibit outstanding optical properties, which allow the application of magneto-optical techniques established in semiconductor spin physics, such as time-resolved Faraday/Kerr rotation~\cite{odenthal2017,belykh2019_nc}, optical spin orientation~\cite{Kopteva2024OO}, optically-detected magnetic resonances~\cite{belykh2022_nl}, spin-flip Raman scattering~\cite{kirstein2022_nc}, spin noise~\cite{Kozlov_2025}, etc. These techniques result in strong spin signals from lead halide perovskite bulk, two-dimensional materials, and nanocrystals. 
Similar to conventional semiconductors, the spectrum of spin-dependent effects in perovskites is largely extended by the hyperfine interaction of the carrier spins with the nuclear spin system~\cite{kirstein2022_am,Kirstein2023_nc,Kotur_2026_DNP}. Due to their low  mobility at cryogenic temperatures, spin signals from both electrons and holes localized in different crystal sites can be traced simultaneously~\cite{kirstein2022_am,Kudlacik_2024OO}. It is also interesting that within the large class of lead halide perovskites of composition $A$Pb$X$$_3$  different possible cations ($A=$MA$^+$, FA$^+$, Cs$^+$) and halogens ($X=$I, Br, Cl) the spin-dependent parameters do not differ greatly despite the large variation of the band gap from 1.5 to 3.2~eV. Currently, spin studies of perovskite semiconductors are still in their initial stage and far from having comprehensive understanding. In particular, although perovskites show relatively long coherent spin precession, extended over tens of nanoseconds, no information about the coherence time $T_2$ of individual spins is available. Indeed, in most of experiments on spin ensembles, the measurement of the $T_2$ time is hindered by the spin dephasing in the ensembles, caused by the spread of $g$-factors and the interactions with nuclear spin fluctuations, resulting in a decay characterized by the spin dephasing time $T_2^*$.

In this manuscript, we study the coherent spin dynamics of electrons and holes in \CFI single crystals by means of time-resolved Faraday rotation. After pulsed excitation, we observe Larmor precession of their spin polarization about an external magnetic field, decaying during the ensemble dephasing time $T_2^*$ amounting up to 21~ns. For much longer delays, up to 1 ms, we find a revival of the spin polarization signal due to synchronization (locking) of the spin precession modes imprinted by periodic laser excitation into the spin ensemble. This spin mode locking (SML) allows us to overcome the spin dephasing caused by ensemble inhomogeneities and to measure the spin coherence times $T_2$ of electrons and holes. Originally, the SML effect was discovered in (In,Ga)As/GaAs quantum dots~\cite{Greilich2006_SML,Greilich2007NF,Yakovlev_Ch6} with the electron $T_2=3$~$\mu$s, in line with the discussion above, and recently observed in CsPb(Cl,Br)$_3$ perovskite nanocrystals with hole $T_2=13$~ns~\cite{kirstein2023_SML}. Here, we report the first observation of the SML effect in bulk semiconductors, unravelling
an extremely long electron spin coherence time $T_2$ at cryogenic temperatures, approaching 1~ms, which is among the longest reported for solid state systems without isotopic purification or dynamic decoupling. 
The long $T_2$ time is accompanied by a similarly long value of the longitudinal spin relaxation time $T_1$. The latter is measured using the resonant spin inertia technique based on optically-detected magnetic resonance (ODMR) with Faraday rotation detection giving  $T_1=0.4$~ms for electrons and $0.15$~ms for holes.


\subsection*{Coherent spin dynamics in \CFI crystal}

\begin{figure}[hbt!]
\begin{center}
\includegraphics[width=1\columnwidth]{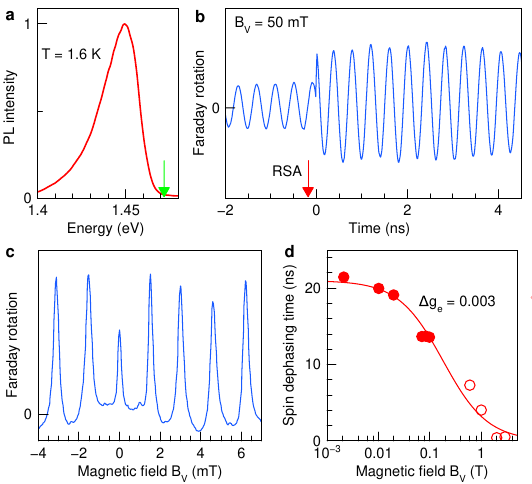}
\caption{\label{fig:Intro} {\bf Coherent spin dynamics of electrons in a \CFI crystal at $T=1.6$~K.} 
(\textbf{a}) Photoluminescence spectrum excited at 3.1~eV photon energy. The photon energy of 1.471~eV used in TRFR is marked by the green arrow. 
(\textbf{b}) Faraday rotation dynamics measured in a Voigt magnetic field of 50~mT.  $T_\text{R}=13.2$~ns. The red arrow indicates the time delay of $\Delta t=-200$~ps between pump and probe pulses used for the RSA measurement in panel C. 
(\textbf{c}) Resonant spin amplification versus magnetic field in Voigt geometry. 
(\textbf{d}) Magnetic field dependence of the electron spin dephasing time $T_{\rm 2,e}^*$ obtained from RSA (closed symbols) and FR dynamics (open symbols). The line shows a fit with eq.~\eqref{eq:InhDeph} using the fitting parameter $\Delta g_{\rm e}=0.003$. 
}
\end{center}
\end{figure}

The PL spectrum of the \CFI crystal at 1.6~K temperature is shown in Fig.~\ref{fig:Intro}a. A specific feature of the lead halide perovskites caused by the low carrier mobility is the presence of resident electrons and holes, which are localized at cryogenic temperatures. Their emission partly overlaps with the exciton emission and is Stokes-shifted to lower energies, resulting in a PL line with 25~meV width. In this sample the PL maximum at 1.45~eV arises from the recombination of spatially-separated, localized electrons and holes, and the exciton emission contributes the high energy flank of the PL line. The exciton energy is at about 1.47~eV, see the detailed study of similar FA$_{0.9}$Cs$_{0.1}$PbI$_{2.8}$Br$_{0.2}$ crystals~\cite{Kopteva2024OO,Kudlacik_2024OO}. 

To study the coherent spin dynamics of resident carriers we use the time-resolved Faraday rotation (TRFR) technique, which has been employed successfully earlier for lead halide perovskite bulk crystals~\cite{belykh2019_nc,kirstein2022_am,kirstein2022_nc,kirstein2022MAPI}, polycrystalline films~\cite{odenthal2017,garcia2021,Grigoryev2021}, and nanocrystals~\cite{Crane2020,Grigoryev2021,Zhu2024review,Meliakov2023_nm,Meliakov2025_paper2}.
The laser photon energy is set to the Faraday rotation (FR) amplitude maximum at 1.471~eV, which about coincides with the exciton energy, see the green arrow in Fig.~\ref{fig:Intro}a.
Spin polarization of resident electrons and holes is created by circularly-polarized pump pulses, and probed by the Faraday rotation of linearly-polarized probe pulses. A magnetic field  ($B_{\rm V}$) is applied in the Voigt geometry (transverse to the optical axis), which leads to Larmor precession of the oriented carrier spins about the magnetic field.

A typical TRFR signal, measured at $T = 1.6$~K in $B_{\rm V} = 50$~mT, is given in Fig.~\ref{fig:Intro}b. The spin dynamics show oscillations with the Larmor precession frequency $\omega_{\rm L,e(h)} = |g_{\rm e(h)}|\mu_B B_{\rm V}/\hbar$, where $g_{\rm e(h)}$ is the electron (hole) $g$-factor, $\mu_B$ is the Bohr magneton, and $\hbar$ is the reduced Planck constant. By fitting the measured dynamics with eq.~\eqref{eq:TRFR}, we evaluate the $g$-factor absolute value of $3.65$. Comparing the dependence  on the bandgap energy~\cite{kirstein2022_nc}, we conclude that this $g$-factor corresponds to an electron and should be positive.

In weak magnetic fields the spin precession amplitude shows a weak decay on a time scale of 5~ns. The spin dephasing times $T_2^*$ cannot be evaluated by fitting the FR dynamics with eq.~\eqref{eq:TRFR}, as the spin precession does not decay sufficiently during the laser repetition period of $T_\text{R}=13.2$~ns, and is still pronounced at the negative time delays in  Fig.~\ref{fig:Intro}b.  In this case, characterized by  $T_2^*>T_\text{R}$, spin accumulation effects under periodic laser excitation have to be considered~\cite{yugova2012,Kirstein2025RSA}. An example is the resonant spin amplification (RSA) effect: Spin precession modes excited by subsequent pump pulses can interfere constructively when phase-synchronized, enhancing the FR signal, while those out of phase interfere destructively~\cite{Awschalom1998,yugova2012}. The phase synchronization condition (PSC) reads
\begin{equation}
\omega_L=\frac{2\pi N}{T_\text{R}} \, ,
\label{eq:PSC}
\end{equation}
where $N$ is an integer number. The effect shows up in the magnetic field dependence of the FR amplitude measured at small negative time delays $|\Delta t| \ll T_\text{R}$, with pronounced maxima at the field strengths where the PSC is fulfilled.

The RSA signal for $T_\text{R}=13.2$~ns measured at $\Delta t=-200$~ps is shown in Fig.~\ref{fig:Intro}c. It comprises a series of sharp peaks separated by $\frac{\hbar\omega_R}{g_{\rm e} \mu_B}=1.5$~mT ($\omega_\text{R} = 2\pi/T_\text{R}$). Their half width at half maximum of $\Delta B = 0.16$~mT in weak magnetic fields corresponds to the spin dephasing time $T_{\rm 2,e}^*=\frac{\hbar}{g_{\rm e} \mu_B \Delta B} \approx 21$~ns. This value is equal to the longest electron spin dephasing time reported for lead halide perovskite crystals, measured in MAPbI$_3$~\cite{Kirstein2025RSA}. 
At $B_{\rm V} \approx 100$~mT we obtain $T_{\rm 2,e}^*\approx 13$~ns, see the complete RSA scan in Fig.~\ref{SI:RSA}. In higher magnetic fields, the decay of the FR dynamics becomes prominent and the spin dephasing time can be evaluated from its fit with eq.~\eqref{eq:TRFR}, see Fig.~\ref{SI:Magn}. The magnetic field dependence of $T_{\rm 2,e}^*$ combining the results from RSA and TRFR is shown in Fig.~\ref{fig:Intro}d. It strongly decreases with growing magnetic field, which is expected for a system with a finite spread  $\Delta g$ of the $g$-factor distribution described by: 
\begin{equation}
\frac{1}{T^*_{\rm 2,e(h)}(B_\text{V})} = \frac{1}{T^*_{\rm 2,e(h)}(0)} + \frac{\Delta g_{\rm e(h)} \mu_B B_\text{V}}{\hbar}.
\label{eq:InhDeph}
\end{equation}
The fit to the data with this equation gives $T^*_{\rm 2,e(h)}(0)=21$~ns and $\Delta g_{\rm e} = 0.003$, see the line in Fig.~\ref{fig:Intro}d.


\begin{figure*}[hbt!]
\begin{center}
\includegraphics[width=1.25\columnwidth]{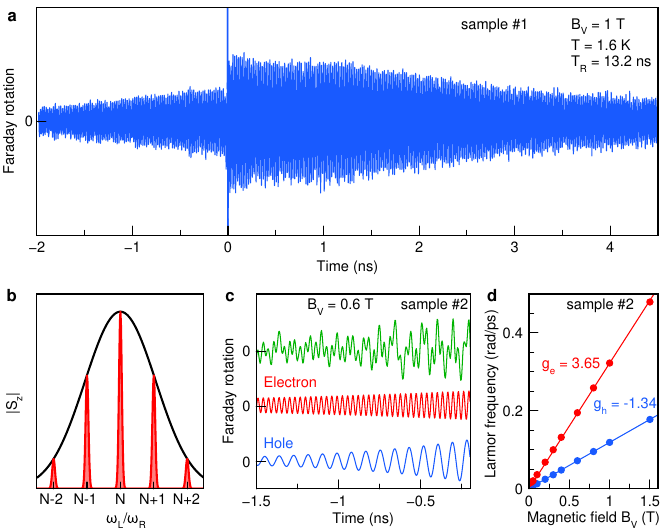}
\caption{\label{fig:SML} \textbf{Spin mode locking of electrons and holes in \CFI crystals at $T=1.6$~K.} 
(\textbf{a}) Faraday rotation dynamics measured in a Voigt magnetic field of 1~T. $T_\text{R}=13.2$~ns. 
(\textbf{b}) Schematic illustration of the spectrum of precession modes in a spin ensemble without (black line) and with SML effect (red line). 
(\textbf{c}) Faraday rotation dynamics measured at negative delays for sample~\#2 (green line), where the SML of holes is pronounced. $B_{\rm V}=0.6$~T. Also, the electron (red line) and hole (blue line) spin dynamics extracted by fitting the experimental dynamics with eq.~\eqref{eq:TRFR} are shown.  
(\textbf{d}) Magnetic field dependence of the electron (red symbols) and hole (blue symbols) Larmor precession frequencies in sample~\#2. The lines are linear fits with slopes corresponding to the electron and hole $g$-factors of $g_{\rm e}=3.65$ and $g_{\rm h}=-1.34$. 
}
\end{center}
\end{figure*}

\subsection*{Spin mode locking of electrons and holes} 

One can see in Fig.~\ref{fig:Intro}d, that the RSA requirement $T_2^*>T_\text{R}$ becomes invalid in magnetic fields stronger than 100~mT, as the spin polarization decays before arrival of the next pump pulse. However, in this case another spin accumulation effect can be observed when the condition $T_2>T_\text{R}$ is fulfilled. This effect is called spin mode locking (SML) of the carrier spin coherences that reveals itself as a continuous echo-like increase of the spin polarization right before the arrival of the next pump pulse (Fig.~\ref{fig:SML}a). It was discovered in singly-charged (In,Ga)As quantum dots~\cite{Greilich2006_SML,Yakovlev_Ch6} and recently it was reported for holes in CsPb(Cl,Br)$_3$ perovskite nanocrystals~\cite{kirstein2023_SML}.

The spectrum of the optically excited Larmor precession modes, whose width is determined by $\Delta g$ and the magnetic field strength, is shown by the black line in  Fig.~\ref{fig:SML}b. Here, the red lines show the modes matching the phase synchronization condition of eq.~\eqref{eq:PSC}. The origin of the SML is similar to that of the RSA~\cite{yugova2012}. In the RSA regime, the spectrum of optically excited modes is narrower than the distance between modes satisfying the PSC. With increasing magnetic field this spectrum broadens (shown by the black solid curve in Fig.~\ref{fig:SML}b) and, therefore, it includes several modes  (red colored peaks in Fig.~\ref{fig:SML}b) satisfying the PSC and enhancing the spin accumulation. The SML signal is the sum of several discrete PSC modes, becoming maximal at the time of pump pulse arrival. As a result, the SML signal decays during the time $T_2^*$ after the pump pulse, but undergoes a revival at negative delays approaching the pump pulse. Note that the characteristic times of amplitude increase at negative delays and amplitude decrease at positive delays are the same and equal to $T_2^*$.

Spin dynamics measured at $B_\text{V} = 1$~T for $T_\text{R}=13.2$~ns in a \CFI crystal (sample~\#1)  are shown in Fig.~\ref{fig:SML}a. The dynamics are characteristic for the SML effect. Here, strong SML signal emerges at negative delays with an amplitude steadily increasing when approaching the next pump pulse at $t=0$, reaching about 50\% of the amplitude at short positive delays. In this sample, the spin dynamics are dominated by the contribution of resident electrons with $T^*_{2,e}=4$~ns, while the hole contribution is small. Sample~\#2 with the same composition shows about equal contributions of electrons and holes to the spin dynamics. One can see in Fig.~\ref{fig:SML}c, that the SML dynamics shown by the green line contains two Larmor frequencies. The respective contributions of electrons and holes extracted by a corresponding fit are shown by the red and blue lines, respectively. The spin dephasing times are 2~ns for the electrons and 1~ns for the holes. The magnetic field dependences of their Larmor precession frequencies and according fits giving the $g$-factors $g_{\rm e}=3.65$ and $g_{\rm h}=-1.34$ are shown in Fig.~\ref{fig:SML}d. Their signs are determined from the universal dependences of the $g$-factors on the band gap~\cite{kirstein2022_nc}. 

The SML amplitude, that we evaluate as the FR amplitude at small negative delays normalized to the FR amplitude at small positive delays, $A_{\rm FR}(-\delta t)/A_{\rm FR}(+\delta t)$, depends on the concentration of the resident carriers, which in turn can vary with the conditions from photocharging. As a result, it varies not only for the two studied samples, but also for different positions of the same sample monitored during the course of measurements. For example, for sample~\#1 with fixed $T_\text{R}=13.2$~ns, the SML amplitude reaches 1 in Figure~\ref{SI:SML}, while it is 0.5 in Fig.~\ref{fig:SML}a and 0.13 in Fig.~\ref{fig:T2}b.

It is important to note that the SML effect has not been observed so far for bulk crystal\addV{s}. Its realization for electrons and holes in \CFI crystals evidences that these resident carriers are sufficiently well localized and their spin coherence times exceed the $T_\text{R}=13.2$~ns used in this experiment. Entering the SML regime gives us access to measurement of the $T_2$ times by all-optical techniques~\cite{Yakovlev_Ch6}.  

\subsection*{Spin coherence times $T_2$ of electrons and holes} 

The SML effect provides an intuitive approach to evaluation of the spin coherence time $T_2$, namely, one can increase the laser repetition period $T_\text{R}$ until the SML signal disappears which occurs, when spin coherence is lost during $T_\text{R}$. The associated $T_\text{R}$ value is an estimate for the $T_2$ time. We use two different laser systems to cover the repetition period range $T_\text{R}$ from 13.2~ns up to 1~ms, see Methods. 

For the laser system operating at 76~MHz the $T_\text{R}$ can be varied by an external pulse picker from 13.2~ns to 33.3~$\mu$s. The power per pump pulse is kept the same, while the average power corresponding to 3~mW at 76~MHz decreases accordingly with the period increase. Most importantly, for the whole range of $T_\text{R}$ the SML amplitude remains about constant evidencing that $T_{2,e}>33.3$~$\mu$s. 

\begin{figure*}[hbt!]
\begin{center}
\includegraphics[width=2\columnwidth]{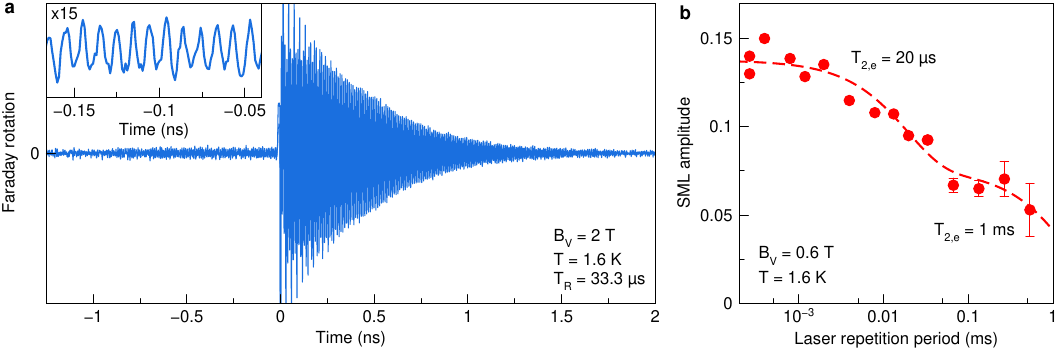}
\caption{\label{fig:T2} \textbf{Dependence of SML on laser repetition period and temperature in \CFI crystal.} 
(\textbf{a}) Faraday rotation dynamics measured with laser system with low repetition rate of 30~kHz ($T_\text{R}=33.3$~$\mu$s).  $B_{\rm V}=2$~T and $T=1.6$~K.  Inset shows zoomed oscillations at negative delays. 
(\textbf{b}) SML amplitude dependence on laser repetition period measured for $T=1.6$~K at $B_{\rm V}=0.6$~T. The data are normalized on the FR amplitude at short positive delays. The dashed line is a fit with sum of two exponents, which gives $T_{2,e}=20$~$\mu$s and 1~ms.   
}
\end{center}
\end{figure*}

For a further increase of $T_\text{R}$ up to 1~ms, we employ the 30~kHz laser system equipped with an internal pulse picker. An example of the FR dynamics measured with this laser at $B_{\rm V}=2$~T for $T_\text{R}=33.3$~$\mu$s is shown in Fig.~\ref{fig:T2}a. At positive delays the dynamics include hole and electron spin precession signals, but at negative delays only oscillations in line with the electron $g$-factor are seen, see the inset in Fig.~\ref{fig:T2}a. A further increase of $T_\text{R}$ leads to a decrease of the SML amplitude that can be followed up to $T_\text{R} = 0.56$~ms. At larger $T_\text{R}$ the decrease of the FR signal intensity results in a signal-to-noise ratio in the measurements too small to reliably detect oscillations at negative delays. 

The dependence of the SML amplitude on $T_\text{R}$ is shown in Fig.~\ref{fig:T2}b. We highlight the logarithmic time scale chosen to present the repetition period covering four orders of magnitude from fractions of a microsecond to a millisecond. The SML amplitude decreases for large periods in two characteristic steps. It can be fitted by the sum of two exponential decays providing estimates for two distinct electron spin coherence times of $T_\text{2,e}=20$~$\mu$s and 1~ms.

With increasing lattice temperature the SML becomes continuously weaker, while $T_2$ stays in the microsecond range at least up to 13~K (see section~S2.3 in the Supplementary Materials). 
We suggest that the temperature decrease of SML amplitude is related to the diminishing of the nuclear spin focusing effect known to strongly increase the SML efficiency in (In,Ga)As quantum dots~\cite{Greilich2007NF,Yakovlev_Ch6} by shuffling spin precession modes that originally do not match to modes fulfilling the PSC. 

The hole SML signal is pronounced at $T=1.6$~K and 6~K in the sample~\#2 for $T_\text{R}=13.2$~ns, see Figs.~\ref{fig:SML}c and \ref{SI:hole6K}, while the mode-locking amplitude is too small for measurements with larger repetition periods. Therefore, we can only conclude that $T_\text{2,h}>T_\text{R}=13.2$~ns.



\subsection*{Longitudinal spin relaxation times $T_1$ of electrons and holes}


We now turn to comparing the transverse to the longitudinal spin relaxation time $T_1$, setting a fundamental limit for the spin coherence time according to $T_2<2T_1$. To measure $T_1$, we use the technique of optically-detected magnetic resonance (ODMR) stimulated by optical pumping, where the FR signal is used for detection~\cite{belykh2022_RSI,belykh2022_nl}. The $T_1$ time is evaluated by the resonant spin inertia approach. In these experiments, the magnetic field is applied in the Faraday geometry (along laser beam) and the laser photon energy set to 1.463~eV. 
A radio frequency ({\it rf}) electromagnetic field with frequency $f_{rf}$ is used to depolarize optically oriented carrier spins, which works mostly efficiently when $f_{rf}$ is in resonance with the carrier Larmor precession frequency. 


\begin{figure}[hbt!]
\begin{center}
\includegraphics[width=1\columnwidth]{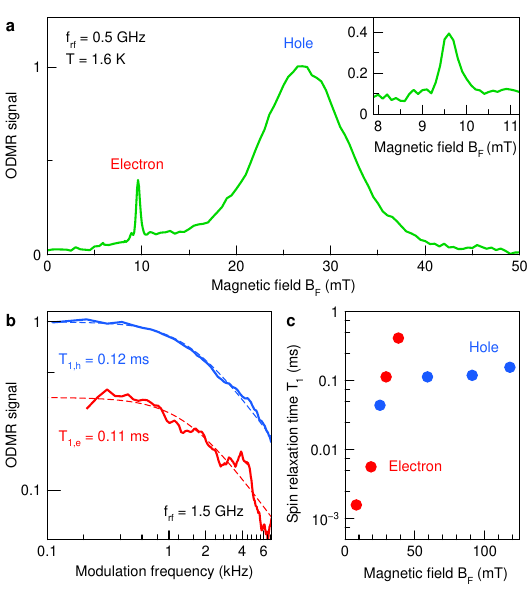}
\caption{\label{fig:ODMR} \textbf{ODMR in \CFI crystal at $T=1.6$~K.}
(\textbf{a}) ODMR signal dependence on the Faraday magnetic field, measured for $f_{rf}=0.5$~GHz and $f_\text{mod}=0.1$~kHz. The insert shows a scan across the low magnetic field peak with smaller field strength steps.
(\textbf{b}) Modulation frequency dependence of the ODMR signal at $f_{rf}=1.5$~GHz for the electrons (red, $B_\text{F}=29$~mT) and the holes (blue, $B_\text{F}=91$~mT). The dashed lines are fits using eq.~\eqref{eq:RSI}.
(\textbf{c}) Magnetic field dependence of the electron (red) and hole (blue) longitudinal spin relaxation times $T_1$. 
}
\end{center}
\end{figure}

Figure~\ref{fig:ODMR}a shows the ODMR signal amplitude as function of the Faraday magnetic field measured for $f_{rf}=0.5$~GHz and $f_\text{mod}=0.1$~kHz. It has two resonances at $B_\text{F}=9.6$ and $27$~mT. The fields of these resonances scale linearly with $f_{rf}$, see Fig.~\ref{SI:g}, giving absolute values of the $g$-factors of 3.6 for the low magnetic field peak and $1.2$ for the high magnetic field peak. These $g$-factors coincide with the electron and hole $g$-factors measured by TRFR and, therefore, we assign them to electrons and holes, respectively. The half widths at half maximum of the ODMR lines correspond to electron and hole spin dephasing times of $T_{\rm 2,e}^*=17$~ns and $T_{\rm 2,h}^*=1.5$~ns. These values are in good agreement with the results of the TRFR experiments.

Modulation of the \textit{rf} field with frequency $f_\text{mod}$ and tracking the dependence of the ODMR amplitude  on $f_\text{mod}$ allow us to extract the spin relaxation time $T_1$~\cite{belykh2022_RSI,belykh2022_nl}. When $f_\text{mod}$ is increased so that the modulation period becomes comparable with or shorter than the $T_1$ time, the accumulated spin polarization (and ODMR amplitude) starts to decrease according to eq.~\eqref{eq:RSI}. Thus, this method allows one to measure $T_1$  selectively for electron and hole by tuning the \textit{rf} field into resonance with the corresponding Larmor precession frequency, which is not possible for the commonly used TRFR spin inertia technique~\cite{Heisterkamp2015,Kirstein2025RSA}. The dependences of the electron and hole ODMR amplitudes on $f_\text{mod}$ for $f_{rf}=1.5$~GHz are shown in Fig.~\ref{fig:ODMR}b. 
Their fits with eq.~\eqref{eq:RSI} provide almost equal electron and hole spin relaxation times of $T_{\rm 1,e}= 0.11$~ms and $T_{\rm 1,h}= 0.12$~ms, respectively.

Figure~\ref{fig:ODMR}c shows the longitudinal spin relaxation times of electrons and holes measured in different magnetic fields. Here, the resonant fields of the electron and hole ODMR lines are achieved by varying $f_{rf}$ in the range from 0.5 to 2~GHz. The $T_{\rm 1,h}$ time weakly depends on magnetic field, increasing from 40~$\mu$s at 27~mT to $0.15$~ms at 117~mT. While, the $T_{\rm 1,e}$ time strongly increases from 2~$\mu$s at 9.6~mT to 0.4~ms at 38~mT. Such a behavior is expected for localized carriers interacting with nuclear spin fluctuations during a finite correlation time~\cite{Smirnov2018,Kudlacik_2024OO}. 
Note that $T_\text{1,e} = 0.4$~ms at $B_\text{F}=38$~mT is comparable to $T_\text{2,e}$, while $T_\text{1,e}$ shows an increase for stronger fields, where $T_\text{2,e}$ is measured. 



\subsection*{Comparison of spin dynamics in solid state systems}



To summarize, three different optical techniques based on the detection of spin signals via Faraday rotation have been used to provide a comprehensive characterization of the electron and hole spin dynamics in \CFI perovskite crystals: resonant spin amplification, spin mode locking, and optically detected magnetic resonance with the resonant spin inertia approach. They allow us to measure the spin dephasing times $T^*_2$, the spin coherence times $T_2$ and the longitudinal spin relaxation times $T_1$. The $T_1$ and $T_2$ for electrons approach one millisecond at the temperature of 1.6~K.  

It is important to note that we report the first observation of the SML effect in bulk semiconductors, namely in \CFI lead halide perovskite crystals. Previously, the SML effect was demonstrated only for electrons and holes strongly confined in (In,Ga)As quantum dots~\cite{Greilich2006_SML,Yakovlev_Ch6} or CsPb(Cl,Br)$_3$ perovskite nanocrystals~\cite{kirstein2023_SML}. Note, that in bulk  perovskite crystals, the SML effect is observed not only in \CFI, but we also find it for holes in MA$_{0.4}$FA$_{0.6}$PbI$_3$ crystals, see Fig.~\ref{SI:MAFA}.




It is instructive to compare the $T_2$ times obtained in  \CFI  crystals with literature data for other semiconductor systems. Among the great variety of available data, see e.g. the review Ref.~\onlinecite{Stano2022}, we limit our comparison to conditions close to our experiment on perovskite semiconductors. Namely, the temperature range of $1.6-4.2$~K accessible with liquid helium cooling, no materials tailoring by isotopic purification, and no use of dynamic decoupling protocols extending the spin coherence. An important factor for high-frequency operation is the possibility for optical processing of the carrier spin coherence. Not all of the materials listed below are suitable for that.  

Among the structures providing quantum confinement for carriers, there are (In,Ga)As quantum dots, where the electron spin $T_{\rm 2,e}$ times measured with SML~\cite{Greilich2006_SML} and Hahn echo decay~\cite{Bechtold2015,stockill2016} amount to $1-3$~$\mu$s. For hole spins confined in (In,Ga)As QDs embedded in microcavity $T_{\rm 2,h}=20$~$\mu$s was obtained in Ref.~\cite{Hogg2025}. In gate-defined GaAs QDs the electron spin $T_{\rm 2,e}=30$~$\mu$s~\cite{bluhm2011}. In silicon-metal-oxide-semiconductor (SiMOS) quantum dots, the electron spin $T_{\rm 2,e}=30$~$\mu$s ~\cite{Huang2024}. 

As indicated, the strong localization of charge carriers at deep impurities or defects provides possibilities for reaching long electron spin coherence times $T_{\rm 2,e}$. However, such systems typically have small oscillator strength for optical transitions and are not suitable for all-optical control. Among them there are: (i) the phosphorous defect in silicon with sub-millisecond times~\cite{Pla2012}, (ii) the nitrogen-vacancy (NV) center in diamond with millisecond times~\cite{takahashi2008,Abobeih2018}, and (iii) the divacancy defects in silicon carbide (SiC) with 1.3~ms~\cite{Miao2020}. Recently, for transition-metal centers (Cr$^{3+}$ and Fe$^{3+}$ ions) in halide double perovskite Cs$_2$NaInCl$_6$, the electron spin $T_{\rm 2,e}=29.5$~$\mu$s and 21.2~$\mu$s at $T=4$~K were measured using pulsed EPR~\cite{Khamkaeo2026}. For an overview these times are collected in Table~\ref{tab:T2_times}.

\begin{table*}
\begin{ruledtabular}
    \centering
    \begin{tabular}{cccccc}
Material  &  Carrier & $T_2$ & Temperature  & Technique  & Reference  \\ \hline
\textbf{lead halide perovskites}&  &  &  &  & \\
 \CFI bulk   & electron & 1 ms & 1.6 K & optical SML & this work\\
\CFI  bulk   & hole & $>13$ ns & 1.6 K & optical SML & this work\\ 
 \MFI  bulk   & hole & $>13$ ns & 1.6 K & optical SML & this work\\ 
CsPb(Cl,Br)$_3$ NCs & hole  & 13 ns  & 1.6 K & optical SML & \onlinecite{kirstein2023_SML}\\  \hline
\textbf{quantum dots}&  &  &  &  & \\
(In,Ga)As QDs & electron  & 3 $\mu$s & 1.6 K & optical SML & \onlinecite{Greilich2006_SML}\\
(In,Ga)As QDs & hole  & 1.6 $\mu$s & 2.1 K & optical SML & \onlinecite{varwig2012}\\
(In,Ga)As QDs & electron  & $1.3-2.7$ $\mu$s & 4.2 K & optical Hahn echo & \onlinecite{Bechtold2015,stockill2016}\\
(In,Ga)As QDs & hole & $20$ $\mu$s & 4 K & optical + EPR & \onlinecite{Hogg2025}\\
GaAs gate-defined  QDs&  electron & $30$~$\mu$s & $\sim0.2$ K$^*$ & EPR Hahn echo  & \onlinecite{bluhm2011}\\
Si/SiO$_2$ QDs& electron  & $30$~$\mu$s  & 1.4 K & EPR Hahn echo & \onlinecite{Huang2024}\\\hline
\textbf{defects \& centers}&  &  &  &  & \\
Si:P& electron & 210~$\mu$s & 0.3 K$^*$ & EPR Hahn echo & \onlinecite{Pla2012}\\
NV in diamond& electron & $20$ ms  & 3.7 K  & optical + EPR & \onlinecite{takahashi2008,Abobeih2018}\\
divacancy in SiC& electron & 1.3~ms & 5 K & optical + EPR  & \onlinecite{Miao2020}\\ \hline
\textbf{centers in perovskites}&  &  &  &  & \\
Cr$^{3+}$ in Cs$_2$NaInCl$_6$& electron & $29.5$~$\mu$s  & 4 K & EPR Hahn echo & \onlinecite{Khamkaeo2026} \\
Fe$^{3+}$ in Cs$_2$NaInCl$_6$& electron & $21.2$~$\mu$s  & 4 K& EPR Hahn echo & \onlinecite{Khamkaeo2026} \\ 
W$^{5+}$ in Ba$_2$CaWO$_{6-\delta}$& electron & $4$~$\mu$s  & 5 K & EPR Hahn echo & \onlinecite{Sinha2019} \\
    \end{tabular}
    \caption{Comparison of the carrier spin coherence times $T_2$ measured in various solid state materials under the following conditions: temperatures in the range of $1.6-5$~K, natural isotope abundance (i.e., no materials tailoring by isotopic purification), and no use of dynamic decoupling protocols for extending the spin coherence. $^*$For these materials we found data only for lower temperatures. 
}
    \label{tab:T2_times}
\end{ruledtabular}
\end{table*}

Thus, the electron spin coherence time reported in this work for perovskites crystals approaching 1 ms is one of the longest in solid state systems with natural abundance and without dynamical decoupling. Combining this feature with the possibility of efficient optical spin generation and manipulation due to the large oscillator strength of exciton and trion transitions, puts lead halide perovskites currently in a pretty unique position for implementing quantum technologies using semiconductors. 
First steps to that end have been taken, for example, by demonstrating all-optical manipulation of the spin coherence in perovskite semiconductors, for example, in CsPbBr$_3$ nanocrystals~\cite{Lin2023}. 
One often demands that 10$^4$ to 10$^6$ operations should be possible within the coherence time for quantum computation. Using ps-light pulses separated by only 10~ps to that end, corresponding to a laser with 100~GHz 
repetition rate (three orders of magnitude higher than the 76~MHz system used here), would require a time window of one to one-hundred microseconds, so far not available, e.g. with III-V quantum dots, but based on our results, likely achievable with perovskites.
For quantum communication, the decisive scale is set by the time it takes to distribute information among the fiber network components. A photon in a fiber needs about 5 microseconds per km. So, for a separation among components of 100~km, the travel time is 500~$\mu$s, so that the quantum information storage time in a quantum repeater has to exceed milliseconds, which seems now within reach.

Also on the material side, further improvements seem feasible: The charge carriers in  lead halide perovskites strongly interact with the lead and halogen nuclei~\cite{kirstein2022_am}. Therefore, isotopic purification might increase the spin coherence time in these materials further. Also, dynamic decoupling might extend the coherence time further, even though it complicates the pulse sequences. In summary, our results represent an important step towards quantum applications, even though still requiring cryogenic conditions.

\section*{Materials and Methods}

\textbf{Samples.}
We studied two samples of \CFI thin crystals: sample~\#1 with thickness of $21$~$\mu$m  and sample~\#2 with thickness of $33$~$\mu$m. Most of the presented data were recorded for  sample~\#1, unless stated otherwise. Data for sample~\#2 were used in Figs.~\ref{fig:SML}c,d. In section S3 we show additional data for a MA$_{0.4}$FA$_{0.6}$PbI$_3$ crystal with a thickness of 30~$\mu$m to demonstrate that the SML effect is not limited to a specific chemical composition of the perovskite. Details of synthesis of these crystals are given in section S1. 

\textbf{Magneto-optical measurements.}
For low-temperature optical measurements, we use a liquid helium cryostat with an insert  for temperatures variable from 1.6\,K up to 300\,K. At $T=1.6$\,K the sample is immersed in superfluid helium, while at $4.2-300$\,K it is held in helium gas. A superconducting vector magnet equipped with three orthogonal pairs of split coils to orient the magnetic field up to 3\,T along any arbitrary direction is used. The magnetic field is denoted as $\textbf{B}_{\rm F}$ when parallel to the laser-light wave vector $\textbf{k}$ (Faraday geometry), and as $\textbf{B}_{\rm V}$ when perpendicular to $\textbf{k}$ (Voigt geometry). 

\textbf{Photoluminescence.}
The photoluminescence (PL) is excited with a continuous wave semiconductor laser emitting with photon energy of 3.1~eV. The spectra are detected by a 0.5\,m monochromator equipped with a silicon charge-coupled-device camera.

\textbf{Time-resolved Faraday rotation (TRFR).}
The coherent spin dynamics are measured using a degenerate pump-probe setup~\cite{Yakovlev_Ch6}. A titanium-sapphire (Ti:Sa) laser generates 1.5\,ps long pulses in the spectral range of $1.265-1.771$\,eV. The laser spectral width is about 1\,nm (about 2\,meV) at the pulse repetition rate of 76\,MHz (repetition period $T_\text{R}=13.2$\,ns). Using an optical pulse picker $T_\text{R}$ can be increased up to 33.3~$\mu$s (repetition rate of 30~kHz). We also use a Light Conversion laser system emitting $1-2$~ps laser pulses at a tunable repetition rate from 1~kHz to 30~kHz (repetition period from $33.3$~$\mu$s to 1~ms), covering photon energies tunable in the range of $0.46-3.93$~eV. The laser output is split into two beams, pump and probe, which pulses can be delayed with respect to one another by a mechanical delay line. The pump and probe beams are modulated using photo-elastic modulators (PEMs) and an optical chopper. The probe beam is linearly polarized and its amplitude is modulated at 84\,kHz. The pump beam is either helicity-modulated at 50\,kHz between $\sigma^+/\sigma^-$ circular polarizations by a PEM or has a fixed circular polarization and is amplitude-modulated by an optical chopper at $100-2000$~Hz. The signal is measured using a lock-in amplifier locked to the difference frequency of the pump and probe modulation. For measuring the spin dynamics induced by the pump pulses, the polarization of the transmitted probe beam is analyzed with respect to its Faraday rotation of the polarization plane using balanced photodiodes. In $B_{\rm V} \neq 0$, the Faraday rotation amplitude oscillates in time, reflecting the Larmor precession of electrons and holes. The dynamics of the Faraday rotation signal can be described by a decaying oscillatory function:
\begin{multline}
A_{\rm FR}(t) = S_{\rm e} \cos (\omega_{\rm L,e} t) \exp(-t/T^*_{\rm 2,e})\\ + S_{\rm h} \cos (\omega_{\rm L,h} t) \exp(-t/T^*_{\rm 2,h}).
\label{eq:TRFR}
\end{multline}
Here $S_\text{e(h)}$ is the electron (hole) amplitude at zero delay, $\omega_{\rm L,e(h)}$ is the electron (hole) Larmor precession frequency, and $T_\text{2,e(h)}^*$ is the electron (hole) spin dephasing time.


\textbf{Optically-detected magnetic resonance with FR detection and resonant spin inertia.}
These methods are described in detail in Refs.~\cite{belykh2022_RSI,belykh2022_nl}. The sample is excited by the emission of a Coherent Chameleon Discovery laser system  emitting 100~fs pulses with a repetition rate of 80~MHz, tunable in a wide spectral range of $1.7-3.1$~eV. Note that for the used technique the fact that the laser is pulsed (rather than continuous wave) is not essential, as long as the laser repetition period of $T_\text{R}=12.5$\,ns is much smaller than the longitudinal spin relaxation time $T_1$. The laser photon energy is tuned close to the exciton resonance of the studied sample.  The polarization of the beam is set to be elliptical. The circular component of the elliptical polarization orients the carrier spins. The spin polarization is detected via the FR of the linearly polarized component. A coil with the diameter of about 1~mm is placed near the sample surface, so that laser beam passed through both the sample and the coil. The coil is used to apply a radiofrequency ({\it rf}) field with frequencies up to 4~GHz. The {\it rf} power is modulated at frequency $f_\text{mod}$ and the corresponding modulation of the spin polarization is detected with a balanced photodetector and a lock-in amplifier.  When the {\it rf} field frequency is in resonance with the carrier Larmor precession frequency, it effectively depolarizes the spin ensemble of optically-oriented carriers. The ODMR signal amplitude ($A_\text{ODMR}$) is the difference of the FR amplitude without {\it rf} field and with {\it rf} field. By scanning the magnetic field applied along the laser beam (Faraday geometry), at the fixed {\it rf} frequency $f_{rf}$ it is possible to measure ODMR spectra and detect the spin resonances for charge carriers in the studied sample. Using this setup it is also possible to measure the longitudinal spin relaxation time $T_1$. Indeed, the ODMR signal does not depend on the {\it rf} modulation frequency $f_\text{mod}$ for low frequencies, where the modulation period is longer than the $T_1$ time. At higher frequencies, $f_\text{mod}\gg 1/T_1$, the signal starts to decrease with growing frequency and the frequency dependence of the signal amplitude allows one to evaluate $T_1$. The dependence of the ODMR signal on $f_\text{mod}$ is given by
\begin{equation}
    A_\text{ODMR}=\frac{A_0}{\sqrt{1+(2\pi T_1 f_\text{mod})^2}}
    \label{eq:RSI}
\end{equation}
Here, $A_0$ is the amplitude in the low modulation frequency limit. In the presented experiments we scan $f_\text{mod}$ in the range $0.1-6$~kHz. This technique is similar to the all-optical spin inertia method~\cite{Heisterkamp2015}. Unlike the latter method, in our case $T_1$ can be measured selectively for a given spin resonance (e.g., of the electron or hole) by setting the {\it rf} frequency or the magnetic field. Therefore, we  call this method the resonant spin inertia technique~\cite{belykh2022_RSI}.

\section*{Supplementary Materials}

This PDF file includes details on the sample synthesis and additional experimental data for the \CFI samples and the MA$_{0.4}$FA$_{0.6}$PbI$_3$ sample. 

\section*{Acknowledgments}

The authors are thankful to N. E. Kopteva and A. Greilich for fruitful discussions. S.R.M, V.V.B., and D.R.Y acknowledge the support of the Deutsche Forschungsgemeinschaft (Project YA 65/28-1, No. 527080192). The work at ETH Z\"urich (B.T. and M.V.K.) was financially supported by the Swiss National Science Foundation (grant agreement 200020E 217589) through the DFG-SNSF bilateral program and by ETH Z\"urich through ETH+ Project SynMatLab.


\textbf{ORCID }\\
Sergey R. Meliakov:  0000-0003-3277-9357 \\
Evgeny A. Zhukov:  0000-0003-0695-0093 \\
Vasilii V. Belykh:  0000-0002-0032-748X  \\
Dmitri R. Yakovlev: 0000-0001-7349-2745  \\
Bekir Turedi:  0000-0003-2208-0737  \\
Maksym V. Kovalenko: 0000-0002-6396-8938  \\
Manfred Bayer:  0000-0002-0893-5949  \\


\section*{Competing interests}
Authors declare no competing interests.

\section*{Data and materials availability.} All data needed to evaluate the conclusions in the paper are present in the paper and/or in the Supplementary Materials.   




\clearpage

\onecolumngrid
\setcounter{page}{1}
\setcounter{section}{0}
\setcounter{figure}{0}
\setcounter{table}{0}
\setcounter{equation}{0}
\renewcommand{\thesection}{S\arabic{section}}
\renewcommand{\thesubsection}{S\arabic{section}.\arabic{subsection}}
\renewcommand{\thepage}{S\arabic{page}}
\renewcommand{\thefigure}{S\arabic{figure}}
\renewcommand{\thetable}{S\arabic{table}}
\renewcommand{\theequation}{S\arabic{equation}}

\renewcommand{\bibnumfmt}[1]{[S#1]}
\renewcommand{\citenumfont}[1]{S#1}


\begin{center}
\textbf{\large Supplementary Materials:}\\[4pt]
\textbf{\large Mode locking of electron and hole spin coherences in perovskite semiconductor crystals}\\[6pt]
Sergey R. Meliakov, Evgeny A. Zhukov, Vasilii V. Belykh, Dmitri R. Yakovlev, Bekir Turedi, Maksym V. Kovalenko, and Manfred Bayer
\end{center}

\section{Sample information}

We experimentally study the electron and hole spin coherence in two samples of thin \CFI crystals: sample~\#1 (technological code CFI6) with a thickness of $21$~$\mu$m  and sample~\#2 (code CFI2) with a thickness of $33$~$\mu$m. Most of the presented data in the text were recorded for sample~\#1, unless otherwise stated. The data shown in Figures~\ref{fig:SML}c and~d corrspond to sample~\#2. Additionally, we study a MA$_{0.4}$FA$_{0.6}$PbI$_3$ crystal with a thickness of $30$~$\mu$m (code MF5).

The crystals were grown using the well-established inverse temperature crystallization method~\cite{S1,S2}. This approach is known to yield high-quality single crystals with well-defined facets and long carrier diffusion lengths~\cite{S3}. \MFI single crystals were synthesized following the reported procedure in Ref.~\cite{S4}. Briefly, a 1.7 M precursor solution in $\gamma$-butyrolactone (GBL) was prepared with a 1:1 molar ratio of organic cations (MAI and FAI) to PbI$_2$. \CFI single crystals were grown using a similar procedure~\cite{S5}. In this case, a 1.7 M precursor solution in GBL was prepared by mixing CsI, FAI, and PbI$_2$ while maintaining an A-site cation-to-Pb ratio of 1. Before crystal growth, the $5 \times 5$ cm$^2$ glasses properly cleaned by acetone, isopropanol and plasma cleaning. Then, a 3 nm polytetrafluoroethylene  (PTFE) layer was deposited by thermal evaporation. The precursor solutions were injected between two PTFE-coated glass substrates and slowly heated from $52^\circ$C to $120^\circ$C at a rate of $2^\circ$C/h. The single crystals studied here exhibit rhombohedral morphologies with lateral dimensions of approximately $2-3$~mm.

\section{Additional experimental data for \CFI crystals}

\subsection{Spin precession and dephasing at different magnetic fields}

\begin{figure}[hbt!]
\centering
\includegraphics[width=0.8\columnwidth]{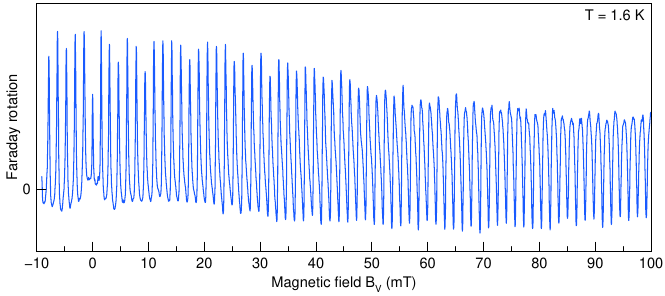}
\caption{
Resonant spin amplification in \CFI crystals for the magnetic field range from $-10$ to $100$~mT. The singal is detected at the time delay of $\Delta t=-200$~ps, $T_\text{R}=13.2$~ns, and $T=1.6$~K. The low field part of this scan is shown in Figure~\ref{fig:Intro}c. The peak widths from this RSA scan are used to determine spin the dephasing time for $B_\text{V}<100$~mT in Figure~\ref{fig:Intro}d.
}
\label{SI:RSA}
\end{figure}

Figure~\ref{SI:RSA} shows the magnetic field dependence of the Faraday rotation signal at the time delay of $t=-200$~ps. The dependence demonstrates resonant spin amplification effect revealed as a series of peaks, at the magnetic fields satisfying the phase synchronization condition with periodic laser pulses. One can see that for increasing magnetic field the peak width increases and corresponding  spin dephasing time $T_2^*$ decreases (see Eq.~(2) in the main text) due to the spread of $g$ factors. This RSA spectrum allows one to extract the spin dephasing times from the peak widths and plot them as a function of magnetic field, see closed symbols in Figure~\ref{fig:Intro}d in the main text. 

\begin{figure}[hbt!]
\centering
\includegraphics[width=0.8\columnwidth]{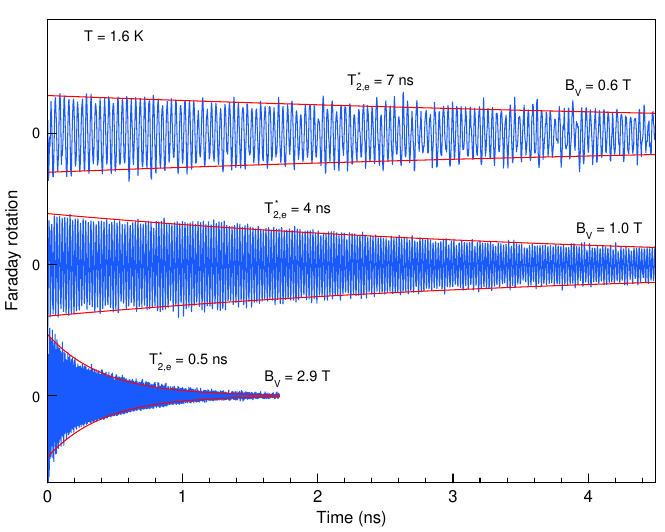}
\caption{
Faraday rotation dynamics measured in different Voigt magnetic fields. The red solid lines show fits with an exponential decay and corresponding spin dephasing times. These times are shown by the open symbols in Figure~\ref{fig:Intro}d.
}
\label{SI:Magn}
\end{figure}

At even higher magnetic fields $T_2^*$ time becomes shorter than the pulse repetition period $T_{\rm R}$ and can be determined directly from the decay of FR dynamics (Fig.~\ref{SI:Magn}). Corresponding times as a function of the magnetic field are shown by open symbols in Figure~\ref{fig:Intro}d in the main text.

\subsection{Spin mode locking with the highest amplitude}

\begin{figure}[hbt!]
\centering
\includegraphics[width=0.8\columnwidth]{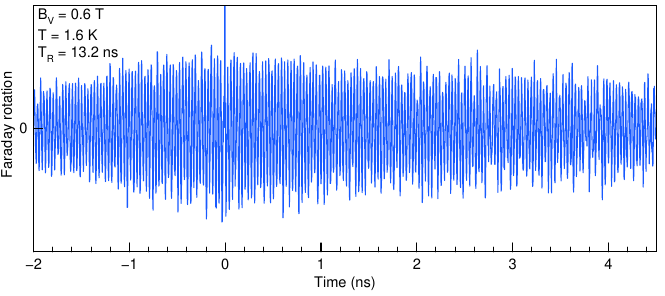}
\caption{
Faraday rotation dynamics measured in a \CFI crystal. $T_\text{R}=13.2$~ns, $B_\text{V}=0.6$~T, and $T=1.6$~K. The ratio between the signal amplitudes at negative and positive delays strongly depends on the position on the sample. Here we show data when these amplitudes are about equal to each other.
}
\label{SI:SML}
\end{figure}

We note that SML amplitude strongly depends on the sample under study and even on the spatial position on a given sample. Figure~\ref{SI:SML} shows FR dynamics with the highest achievable SML amplitude approaching 1.

\subsection{Temperature dependence of the spin mode locking}

\begin{figure}[hbt!]
\centering
\includegraphics[width=0.75\columnwidth]{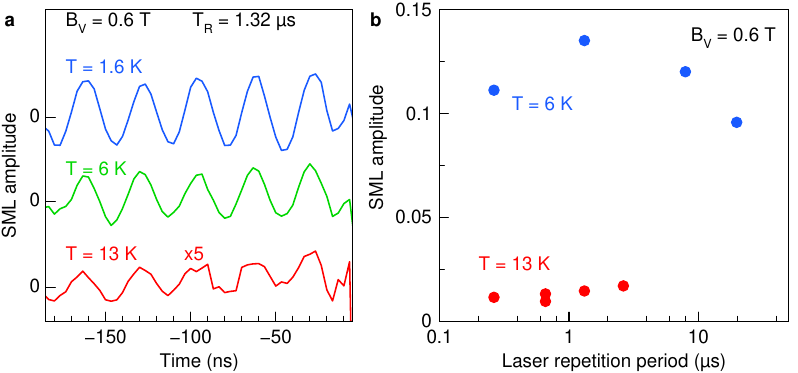}
\caption{Temperature dependence of SML in \CFI crystal. 
(\textbf{a}) SML signals at negative delays normalized on the FR amplitude at positive delays, measured at temperatures of 1.6~K (blue), 6~K (green) and 13~K (red). The 13~K signal is multiplied by a factor of 5. $B_{\rm V}=0.6$~T and $T_\text{R}=1.32$~$\mu$s. 
(\textbf{b}) Dependencies of the normalized SML amplitude on the laser repetition period for temperatures of 6~K (blue) and 13~K (red). $B_\text{V}=0.6$~T. The SML amplitude is about independent of $T_\text{R}$ in these temporal ranges at both temperatures. For longer $T_\text{R}$ the signal-to-noise ratio limits the reliable evaluation of the SML amplitude. From these data we estimate $T_\text{2,e}>20$~$\mu$s at $T=6$~K and $T_\text{2,e}>3$~$\mu$s at $T=13$~K.
}
\label{SI:Temp}
\end{figure}

With increasing lattice temperature the SML becomes continuously weaker, as can be seen in Fig.~\ref{SI:Temp}a where the FR dynamics at negative delays are shown for $T_\text{R}=1.32$~$\mu$s with the temperature increasing from 1.6~K to 13~K. In this range, the SML amplitude decreases from 0.15 at $T=1.6$~K to 0.015 at 13~K. The dependences of the SML amplitude on $T_\text{R}$ at elevated temperatures are shown in Fig.~\ref{SI:Temp}b. They demonstrate that in turn, the SML amplitude is almost independent of the repetition period for $T_\text{R}<20$~$\mu$s at $T=6$~K and $T_\text{R}<3$~$\mu$s at $T=13$~K. Therefore, $T_{\rm 2,e}$ is  still larger than 20~$\mu$s at $T=6$~K and larger than 3~$\mu$s at $T=13$~K. Note that these times are larger than the $T_\text{R}=1.32$~$\mu$s used in Fig.~\ref{SI:Temp}a, and hence the shortening of the $T_2$ time is not the main reason for the temperature decrease of the SML amplitude. We suggest that this decrease is rather related to the diminishing of the nuclear spin focusing effect known to strongly increase the SML efficiency in (In,Ga)As quantum dots~\cite{si_Greilich2007NF,si_Yakovlev_Ch6} by shuffling spin precession modes that originally do not match to modes fulfilling the PSC. For lead halide perovskite crystals, strong effects of the hyperfine interaction of electrons and holes with the nuclear spin bath have been documented in time-resolved coherent spin dynamics~\cite{si_kirstein2022_am} and polarized photoluminescence~\cite{si_Kudlacik_2024OO,si_Kotur_2026_DNP}. 

\label{sec:SI:Temp}

\subsection{Hole spin mode locking at 6~K}
\begin{figure}[hbt!]
\centering
\includegraphics[width=0.75\columnwidth]{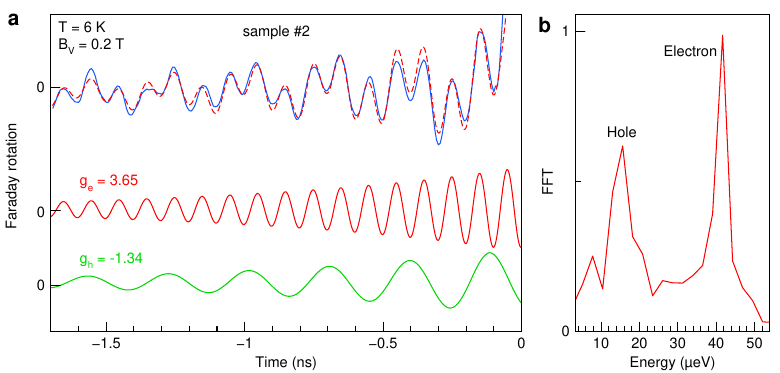}
\caption{
(\textbf{a}) Faraday rotation dynamics measured on the sample~\#2 \CFI crystal. $T_\text{R}=13.2$~ns, $B_\text{V}=0.2$~T, and $T=6$~K. Electron and hole Larmor precession are observed in the SML signal. The red dashed line is a fit with eq.~\eqref{eq:TRFR} using $g_{\rm e}=3.65$ and $g_{\rm h}=-1.34$. The red and green solid lines show the electron and hole spin components, respectively, in the FR dynamics after decomposition.
(\textbf{b}) Fast Fourier transform of the dynamics from panel (\textbf{a}). The lower energy and high energy peaks correspond to the hole and electron Zeeman splittings. 
}
\label{SI:hole6K}
\end{figure}

We observe both electron and hole SML in the sample \#2 with comparable amplitudes at 1.6~K (Fig. 2c in the main text) and also at 6~K (Fig.~\ref{SI:hole6K}).

\subsection{Measurement of electron and hole $g$ factors from ODMR}

\begin{figure}[hbt!]
\centering
\includegraphics[width=0.5\columnwidth]{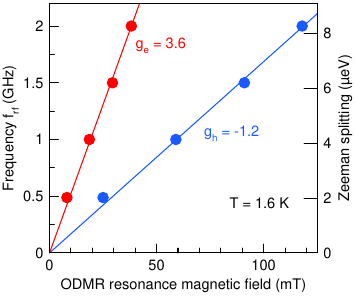}
\caption{  
Magnetic field dependence of the electron (red symbols) and the hole (blue symbols) magnetic resonance frequencies (Zeeman energy splittings) evaluated from the ODMR data at $T=1.6$~K. 
Linear fits to these data (lines) give electron and hole $g$-factors of $g_{\rm e}=3.6\pm0.1$ and $g_{\rm h}=-1.2\pm0.1$. Note that in these experiments only the absolute values of the $g$-factors can be evaluated. Their signs have been based on the basis  of their identification as electron and hole signals. 
}
\label{SI:g}
\end{figure}

We plot the \textit{rf} frequency at which ODMR spectra (Fig. 4a in the main text) are recorded versus the magnetic field for which electron and hole resonances are observed in Fig.~\ref{SI:g}. The dependences for both resonances are close to linear with the slopes corresponding to $|g|$ of 3.6 and 1.2.  These values are close to the absolute values of electron and hole  $g$ factors measured in the pump-probe experiments (Fig.~2d in the main text).

\clearpage

\section{Additional experimental data for MA$_{0.4}$FA$_{0.6}$PbI$_3$ crystals}

\begin{figure}[hbt!]
\centering
\includegraphics[width=1\columnwidth]{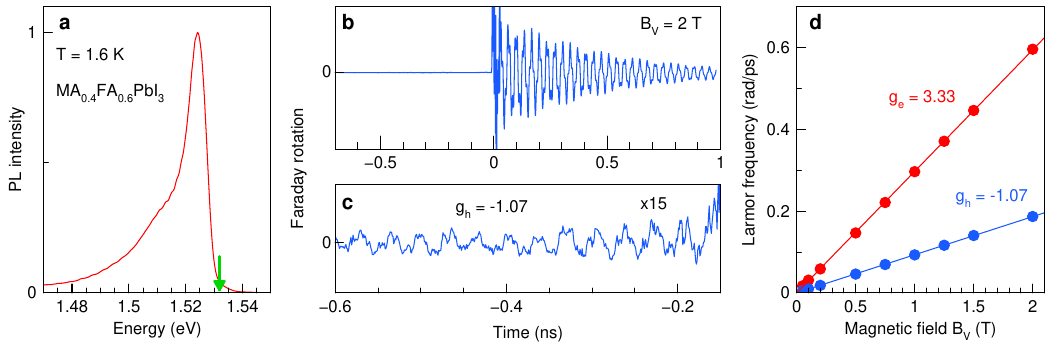}
\caption{
\textbf{Coherent spin dynamics of carriers in a \MFI crystal at $T=5$~K.} 
(\textbf{a}) Photoluminescence spectrum excited using 3.1~eV photon energy, measured at $T=1.6$~K. The green arrow indicates the laser photon energy of 1.532~eV used in the TRFR experiments.  
(\textbf{b}) Faraday rotation dynamics measured at $B_\text{V}=2$~T. The pump and probe powers are 3~mW and $0.5$~mW, respectively. $T_\text{R}=13.2$~ns, and $T=5$~K.
(\textbf{c}) Faraday rotation dynamics at negative delays multiplied with a factor 15, which corresponds to hole SML with the g-factor $g_{\rm h}=-1.07$ for $T_{\rm 2,h}>T_{\rm R}=13.2$~ns. 
(\textbf{d}) Magnetic field dependences of the electron (red) and hole (blue) Larmor frequencies. The solid lines show linear fits with slopes resulting in $g_\text{e}=3.33$ and $g_\text{h}=-1.07$.
}
\label{SI:MAFA}
\end{figure}

We measure the coherent spin dynamics of electrons and holes in an \MFI crystal with a thickness of 30~$\mu$m at $T=5$~K.  Figure~\ref{SI:MAFA}a shows the PL spectrum of this sample at $T=1.6$~K. It has a maximum at 1.52~eV and its shape is similar to the one measured for the \CFI sample, compare with Fig.~\ref{fig:Intro}a. In the TRFR experiments we use a laser photon energy of 1.532~eV. The FR dynamics measured at $B_\text{V}=2$~T are shown in Fig.~\ref{SI:MAFA}b. At positive delays, a FR signal with two Larmor frequencies is observed. The slopes of their dependences on magnetic field (Fig.~\ref{SI:MAFA}d) correspond to electron and hole $g$-factors of $g_\text{e}=3.33$ and $g_\text{h}=-1.07$. At negative delays, the signal amplitude is quite small and we show it more clearly by zooming into Fig.~\ref{SI:MAFA}c.  It increases when approaching zero time delay, i.e. the pump laser arrival time, which is inherent to the SML signals. Here, only oscillations with the hole $g$-factor $g_\text{h}=-1.07$ are seen. Therefore, we conclude that we detect in this sample the SML effect for resident holes, whose spin dephasing time exceeds $T_{\rm R}=13.2$~ns.
This result also demonstrates that the SML can be observed in lead halide perovskite crystals with different compositions. More information about the spin dynamics in MA$_{x}$FA$_{1-x}$PbI$_3$ crystals can be found in Ref.~\cite{si_Gribakin2026}. 

\end{document}